\documentclass[12pt]{article}

\usepackage{amsmath}
\usepackage{amssymb}
\usepackage{verbatim}

\makeatletter

   \@addtoreset{equation}{section}
\makeatother

\newcommand{\be}{\begin{equation}}
\newcommand{\ee}{\end{equation}}
\newcommand{\ba}{\begin{eqnarray}}
\newcommand{\ea}{\end{eqnarray}}

\begin{document}

\vskip 12mm

\begin{center} 
{\Large \bf  Super-spectral curve of  irregular conformal blocks}
\vskip 10mm
{ \large  Dmitri Polyakov$^{a,b,}$\footnote{email:polyakov@scu.edu.cn;polyakov@sogang.ac.kr}
and  Chaiho Rim$^{c,}$\footnote{email:rimpine@sogang.ac.kr} }

\vskip 8mm
$^{a}$ {\it  Center for Theoretical Physics, College of Physical Science and Technology}\\
{\it  Sichuan University, Chengdu 6100064, China}\\
\vskip 2mm

$^{b}$ {\it Institute of Information Transmission Problems (IITP)}\\
{\it  Bolshoi Karetny per. 19/1, Moscow 127994, Russia}\\
\vskip 2mm
$^{c}$ {\it Department of Physics and Center for Quantum Spacetime}\\
{\it Sogang University, Seoul 121-742, Korea}

\end{center}

\vskip 15mm

\begin{abstract}
We use super-spectral curve
to  investigate irregular conformal states
of integer and half-odd integer rank. 
The spectral curve is the loop equation of  
supersymmetrized  irregular matrix model.
The case of integer rank corresponds to the  colliding limit of 
supersymmetric vertex operators of NS sector 
and half-odd integer  to the Ramond sectors.
The spectral curve
is simply integrable at Nekrasov-Shatashvili  limit 
and the partition function 
(inner product of irregular conformal state) 
is obtained from the superconformal structure 
manifest in the spectral curve. 
We present some  explicit forms of the partition function
of integer (NS sector)  and  of half-odd ranks (Ramond sector).  
\end{abstract}

\vskip 12mm

\setcounter{footnote}{0}

\section{\bf Introduction} 
Irregular conformal state is a conformal state, 
but is not a primary or descendent state. 
Rather it is similar to a coherent state
since it is a simultaneous eigenstate of some of positive mode of 
conformal generator. 
The simplest irregular state is the eigenstate of Virasoro $L_{+1}$ mode
which is called Gaiotto state \cite{G_2009}
or Whittaker state \cite{Whittaker}. 
More irregular states have been systematically investigated
for Virasoro and W-irregular state 
\cite{MMM_2009,BMT_2011,KMST_2013,CRZ_2014,CRZ_2015}.

The irregular state is termed as rank $n$ if it is the 
simultaneous eigenstate of Virasoro generators $L_k$ with $n \le k \le 2n$
or of $ W^{(q)}$  generators $W^{ (q)}_k$ with $(q-1)n \le k \le qn$ with  spin $q$. 
However, the construction of the irregular state is not easy to find
because the eigenvalues are not enough to define the state of rank $n\ge 2$. 
One needs additional information how the irregular state behaves 
when all other positive generators applied on the irregular state
such as $W^{ (q)}_k $ with $0 \le k <(q-1)n $. 

The progress is achieved according to  AGT \cite{AGT_2010} 
and the idea of colliding limit \cite{EM_2009,GT_2012}.
 AGT connects Nekrasov partition function 
of N=2 super Yang-Mills theory in 4 dimension with the Liouville conformal block 
in 2 dimension. Colliding limit of Liouville conformal block describes 
the irregular state and in turn closely related with Argyres-Douglas theory 
of N=2 super Yang-Mills theory. 
Colliding limit of the Liouville conformal block is easily investigated 
in terms of irregular matrix model.  
Originally, Penner-type matrix model is suggested from the Liouville conformal block
\cite{DV_2009,EM_2009,IO_2010}
and colliding limit of the Penner-type matrix model results in the irregular matrix model. 

The irregular matrix model is successful to describe irregular states 
and their inner product. 
The partition function is related with the inner product between 
primary state and irregular state or between two irregular states 
depending on the potential of the irregular matrix model. 
However, the success is limited to the case of irregular states of integer rank. 
Virasoro irregular state of integer rank $n$ 
has  eigenvalue of the highest Virasoro generator $L_{2n}$. 
Question arises. Can we find irregular state with half-odd rank, 
that is, irregular state of highest Virasoro generator $L_{2n-1}$? 
The state of rank 1/2 is easily found from the rank 1 if one   limits 
the eigenvalue of $L_2$ vanish. 
However, this trick does not work for rank greater than 1 since this limit 
does not exist since other eigenvalues diverges unless special limit 
is achieved so that the state is a simultaneous eigenstate of $L_1$ and $L_{2n-1}$ only \cite{BMT_2011}.

In this paper, we will present the irregular matrix model of half-odd integer rank
using supersymmetrizing the theory. 
The irregular vertex operator is constructed similar 
to the regular vertex operator \cite{PR_IV_2016,NS_2010,GF_2014}
and is supersymmetrized in  \cite{PR_2016}.
It is noted that the irregular vertex operator with 
half-odd rank appears naturally with Ramond sector 
in the super-symmetrized version. 
This operator is useful for the free field formalism. 
If one includes the screening operators, then one can 
investigate the interacting system of irregular states. 

This paper is organized as following. 
In section 2, we present irregular super-matrix model and its loop equation. 
The matrix model is related with the N=1 super Liouville conformal block and its colliding limit. The loop equation is simply integrable at Nekrasov-Shatashvili limit (NS limit), which is called super-spectral curve.  
In section 3, we consider   irregular states with integer rank. This state is obtained
from the NS sector. Using the super-spectral curve we obtain partition function 
and present the explicit form of rank 1. 
In section 4, irregular states with half-odd rank are considered. 
Partition functions of rank 1/2 and 3/2 are presented. 
In section 5, we present an idea on RG flow equation
corresponding to the operator algebra of the irregular vertices 
from the string field theory. 
Section 6 is the conclusion and discussion.  
Super-spectral curve of the irregular matrix model is presented in the appendix.  
 
\section{Irregular super-matrix model and its spectral curve}
\label{sec_2}
Super-vertex operator  $V_\alpha (\zeta)$ in the NS sector 
is considered in the super-field formalism
\be
V_\alpha (\zeta) = e^{\alpha \Phi} (\zeta)
\ee
where  $\zeta=(z, \theta)$ is the holographic super-coordinate,
$\Phi$ is the super-field and $\alpha$ is the Liouville momentum.
Two point correlation of the vertex operator is normalized as 
in \cite{Rash_Stanish_1996}
\be
\left\langle V_{\alpha_1} (\zeta_1) V_{\alpha_2} (\zeta_2)  \right\rangle 
= (z_{12} -\theta_1 \theta_2)^{-\alpha_1 \alpha_2} 
\ee
where $z_{zb} =z_a-z_b$. 
To find the multi-point correlation in the super Liouville formalism 
one may use  screening operator $V_b (\zeta)$
 in the  presence of background charge $Q=b+ 1/b$.
Primary operator has the  conformal dimension $\Delta_{\alpha} =\alpha(Q-\alpha)/2$ 
and the superconformal system has  central charge $c= 3/2(1 + 2 Q^2)$. 

Explicitly, ($n+2$)-point holomorphic correlation can be calculated  
in the presence of $N$-screening operators
$V_b(\zeta)$ and be put into Selberg integrals
\be
\left\langle 
\prod_{A=1}^{n+2} V_{\alpha_A}(\zeta_A) 
 \right\rangle 
= \int \left [ \prod_{I=1}^N dz_I d\theta_I \right]
\prod_{I<J} (z_{IJ} - \theta_I \theta_J )^{-b^2} 
~ \prod_{I, A} (z_{IA}- \theta_I \theta_A)^{-b \alpha_I} 
\ee
where neutrality condition $ \sum_I \alpha_I + N b =Q$ is assumed. 

To formulate this integral in terms matrix model, we put $ (n+2)$ external operator contribution $  (z_{IA}- \theta_I \theta_A)^{-b \alpha_I} $  into an exponential of a  super-potential  
$ V(\zeta_I) = \sum_A \hat{\alpha}_I \ln  (z_{IA}- \theta_I \theta_A) $ 
with $\hat \alpha = \hbar \alpha$;
\be
{\cal Z}_n
= \int \left [ \prod_{I=1}^N dz_I d\theta_I \right]
\prod_{I<J} (z_{IJ} - \theta_I \theta_J )^{\beta}
~e^{\frac{\sqrt{\beta} } g   \sum_{I}  V(\zeta_I)} \,.
\label{Z_n}
\ee 
This will be called deformed super Penner-type  matrix model. 
Here $\beta =-b^2$ is used instead of $b$.
In addition, $g= i \hbar $ is introduced for later convenience.  
In terms of the new notations, $b=i \sqrt{\beta}$,
$Q=i (\sqrt{\beta} - 1/\sqrt{\beta})$ and 
$\hbar Q = g (\sqrt{\beta} - 1/\sqrt{\beta})$. 

If one applies the colliding limit by fusing $n$ operators to the one at origin
and let  the rest to the infinity after accordingly normalizing 
the partition function at infinity, one 
obtains  a new super potential of the form
$V (\zeta_I) = V_B(z_I) + \theta_IV_F(z_I)$:
$ V_B (z)$ and $V_F(z)$ are bosonic and fermionic part of super-potential.

The loop equation provides the super-spectral curve 
with the deformed parameter $\epsilon$.
(Its derivation is found in Appendix \ref{append_A}).
\begin{align}
x_B (z) x_F (z) + \epsilon  x_F '(z)  = F_F (z)
\label{spectral_curve-F}
 \\
x_B(z)^2 + \epsilon x_B'(z) + x_F(z) V_F' (z) - x_F'(z)  V_F(z) = 2F_B (z)  
\label{spectral_curve-B}
\end{align}
where $ x_F (z)$ ($x_B (z)$)  is anti-commuting (commuting)
 one-point  resolvent $ \omega_F(z)$ ($ \omega_B(z)$)
shifted by potential term,  $x_F(z) = \omega_B(z) - V_F(z) $
( $ x_B(z) = \omega_B(z) + V_B'(z)$). 
 $F_F$ ($F_B$) is also  anti-commuting (commuting) holomorphic function
and  represent  spin 3/2 supercurrent (spin 2 Virasoro) symmetry of the partition function.  
 
Explicitly, the potential obtained from the colliding limit of  
($n+2$) number of NS sector of N=1  super Liouville vertex operators
 is  of the form
\begin{align}
V_B(z_I) & = c_0  \ln  (z_I) - \sum_{k =1}^{n} \frac{c_k} {k  z_I ^k} 
\label{V_B}
\\
V_F(z_I) &=  - \sum_{k =0}^{n} \frac{\xi_k}{z_I^{k+1}} .
\label{V_F}
\end{align}
where  $V_B(z_I)$ is the bosonic part and  $V_F(z_I)$
the fermionic part. 
$ c_k   $ is a   commuting variable defined as 
$c_k= \sum_A \hat \alpha _A z_A^k$
and   $ \xi_k $ is an  anti- commuting variable 
defined as $\xi_k
 = \sum_A \hat \alpha _A  z_A^k \theta_A $. 
The partition function with the new super potential will be called irregular super-matrix model of integer rank $n$. 

It is noted that the matrix model is closely related with 
irregular vertex operator was investigated in \cite{PR_2016}
\be
W_{n}=e^{\sum_{k=0}^{2n}{ \gamma}_kD_\theta^k{ \Phi} (z,\theta)}
\label{ns_vertex}
\ee
where $D_\theta=\theta\partial_z+\partial_\theta$.
$\gamma_k $ is commuting (anti-commuting) when $k$ is even (odd). 
The same potentials $V_B(z_I) $ and $V_F(z_I)$ 
in \eqref{V_B} and \eqref{V_F} are obtained if one contracts
$W_{n}$ with $N$ screening operators $V_b(\zeta)$.
 
There is a slightly different form of the super-matrix model 
due to  Ramond sector. If one uses the vertex operator of Ramond sector 
\be
W_{n-{1\over2}}=e^{\sum_{k=0}^{2n-1}{ \gamma}_kD_\theta^k{ \Phi} (z,\theta)},
\label{ramond_vertex}
\ee
and contracts $W_{n-{1\over2}}$ with $N$ screening operators
$V_b(\zeta)$,
one obtains the irregular matrix model of Ramond sector.
The resulting irregular potential is the one 
similar to \eqref{V_B} and \eqref{V_F}:
\begin{align}
V_B(z_I) & = c_0  \ln  (z_I) - \sum_{k =1}^{n} \frac{c_k} {k  z_I ^k} 
\label{ramond_V_B}
\\
V_F(z_I) &=  - \sum_{k =0}^{n-1} \frac{\xi_k}{z_I^{k+1}} .
\label{ramond_V_F}
\end{align}
The difference from the model of rank $n$ (NS sector) 
is that the commuting variable $c_k$ has unusual constraints.
$c_k$ contains the product of two anti-commuting variables
so that $c_n^2=0 =c_n \xi_{n-1}$. 
This model is called irregular super-matrix model of half-odd rank $(n-1/2)$.  

\section{Partition function of integer rank} 
\label{sec_3}

For the integer rank $n$, the potential is given in \eqref{V_B} and \eqref{V_F}: 
\be
V_B(z) = c_0 \ln (z) - \sum_{k=1}^n \frac{c_k}{k z^k}, ~~~ 
V_F(z) = - \sum_{k=1}^n \frac{\xi_k}{ z^{k+1}}.
\label{potential_rank_n}
\ee 
Here  $c_k$ ($\xi_k $)  is a  commuting (anti-commuting ) variable.
We are using the super- spectral curve 
\eqref{spectral_curve-F} and \eqref{spectral_curve-B} 
using the explicit form of $F_F$ ($F_B$) using 
the potential \eqref{potential_rank_n}. 
\begin{align}
 F_F(z) &=
  \sum_{r=1/2}^{2n-1/2} \frac{ \Omega_{r}} {z^{3/2+r}} 
+ \sum_{r=1/2}^{n-1/2} \frac{ \eta_{r}} {z^{3/2+r}}  \\
F_B(z) & =\sum_{m=0}^{2n} \frac{\Lambda_{m}} {z^{2+m}} 
+  \sum_{m=0}^{n} \frac{d_{m}} {z^{2+m}}.
\end{align}
$\Omega_r $ is anti-commuting number and is defined as   $\Omega_r = \sum_k c_k \xi_{r -1/2-k} - \epsilon ( \delta_{r, 1/2} -(r+1/2)) \xi_{r-1/2} $. 
On the other hand,   $\Lambda_m$ is commuting number, 
$\Lambda_m =  \sum_{k+l =m} c_k c_l/2   - \epsilon (m+1) c_m /2 $. 
It is noted in the appendix  \ref{append_A}
that $\eta_r$  ($d_m$) is an expectation value $\eta_r =g_r (-\hbar^2 \ln Z) $
with  supercurrent $g_r$
($d_m = \ell_m (- \hbar^2 \ln Z) $   
with Virasoro current $\ell_m$).  

This expectation value is the basic tool to find the partition function 
from the super-spectral curves \eqref{spectral_curve-F} 
and  \eqref{spectral_curve-B} 
as noted in bosonic cases
 \cite{NR_2012,CR_2013,CR_2015}. 
The super-flow equation \eqref{etar_gr} and \eqref{dm_ellm} is essential 
to find the moments $d_m$ and $\eta_r$. 
In the following we provide an explicit calculation 
for the simplest case (rank 1). 

For the rank 1, there are 3 flow equations: 2 bosonic  and 
one fermionic:
\begin{align}
d_0 = & \left( c_1 \frac{\partial}{\partial c_1 } + \frac12  \xi_0 \frac{\partial}{\partial \xi_0 }
+\frac32  \xi_1 \frac{\partial}{\partial \xi_1 } \right)  (-\hbar^2 \ln Z)
\label{flow_d0}\\
\eta_{1/2} =&  \left( \xi_1 \frac{\partial}{\partial c_1 }  -c_1  \frac{\partial}{\partial \xi_0 } \right)
(-\hbar^2 \ln Z) + \epsilon \xi_0
\label{flow_eta1} \\
d_1  = &\xi_1  \frac{\partial}{\partial \xi_0  }  (-\hbar^2 \ln Z).
\label{flow_d1}
\end{align}

To solve the flow equations, we need to find the moments
$d_0$, $\eta_{1/2}$ and $d_1$ 
in the left hand side of the flow equations 
as the functional dependence 
of variables $c_1, \xi_1$ and $\xi_2$
from the super-spectral curve. 
The moment $d_0$  is easily identified if one considers 
the dominant contribution of 
the bosonic spectral curve \eqref{spectral_curve-B} 
at large $z$ limit. 
\be
d_0 =  \epsilon  N \left(c_0 + \frac{\epsilon (N-1)}2  \right).
\label{d_0} 
\ee
Therefore, the bosonic flow equation  \eqref{flow_d0}  requires 
the partition function to be of the form 
\be
-\hbar^2 \ln Z = d_0 \log c_1 + A \xi_0 \xi_1/c_1^2 +C.
\label{lnZ_try1}
\ee
Here, $ \xi_0 \xi_1/c_1^2 $ is the homogeneous solution and 
C is a constant independent of $c_1, \xi_1$ and $\xi_2$, 
which can be normalized to be 0. 

The fermionic moment  $\eta_{1/2}$ is obtained if 
we use the large $z$ expansion of  \eqref{spectral_curve-F}:
\be
\eta_{1/2} =\epsilon  N \xi_0
+ \epsilon N_F  (\epsilon (N-1)  + c_0 )  
\label{eta_1/2} 
\ee
where  $N_F=\langle \sum_I \theta_I \rangle$.  
To get the information on $N_F$, we use the fact 
that  $d_m$ and $\eta_r$  should obey  the consistency condition 
due to commutation relations \eqref{super_alg} between generators.
Note that $ [l_m, g_r] = (r-m/2) g_{r+m}$. This requires   
\be
l_m ( \eta_r ) - g_r (d_m ) = (r-\frac m2 ) \eta_{r+m},
\ee
Therefore,  $d_0$ in \eqref{d_0} and $\eta_{1/2} $ in \eqref{eta_1/2} 
has the relation:
$l_ 0 ( \eta_{1/2} )  = (1/2 ) \eta_{1/2}$ 
since  $ g_{1/2}  (d_0 ) = 0$.
This shows that $ \eta_{1/2}$ behaves as  the primary of conformal 
dimension $1/2$. 
There are two anti-commuting variables $\xi_0$ and $ \xi_1/c_1$  of 
dimension $1/2$.  This shows that  
$N_F$  should be proportional to either $\xi_0$ or $ \xi_1/c_1$.
Fermionic filling fraction is anti-commuting and 
is concentrated at $\xi_0$ or $ \xi_1$.
Putting $N_F= N_1 \xi_0  + N_2 \xi_1/c_1 $  
with commuting numbers  $N_1$ and $N_2$, one has 
\be
\eta_{1/2} =\xi_0 \Big( \epsilon  N 
+ \epsilon N_1   (c_0 + \epsilon (N-1)  ) \Big)
+\left( \frac{\xi_1} {c_1} \right) \Big( c_0 +N_2\epsilon (N-1) \Big). 
\label{eta_1/2_try1}
\ee

The flow equation  \eqref{flow_eta1} together with 
\eqref{lnZ_try1} and  \eqref{eta_1/2_try1} is rewritten as 
\be
\xi_0 \Big( \epsilon  N 
+ \epsilon N_1   (c_0 + \epsilon (N-1)  ) \Big)
+\left( \frac{\xi_1} {c_1} \right) \Big( c_0 +N_2\epsilon (N-1) \Big)
= \xi_0  \epsilon+ \left(  \frac{ \xi_1}{c_1} \right) ( d_0 - A ) 
\ee 
Therefore, the flow equation reduces to algebraic identities: 
\begin{align}
& \epsilon  N 
+ \epsilon N_1   ( c_0 +\epsilon (N-1)  )= \epsilon 
\nonumber\\
&N_2 ( c_0 + \epsilon (N-1)  )=  d_0 - A 
\end{align}
which fixes $N_1$ and $A$ as a function of $c_0, N$ and $N_2$:
\be
 N_1= - \frac{N-1}{ c_0+ \epsilon (N-1)  } ,~~~
A=d_0 - N_2 ( c_0 +\epsilon (N-1) ).
\ee

As a result, the partition function is given as 
\be
-\hbar^2 \ln Z
= d_0  \ln c_1 +  \left(  \frac{   \xi_0 \xi_1}{c_1^2 } \right)  
( d_0  - N_2  ( c_0 +\epsilon (N-1) ).
\label{lnZ_final}
\ee

Finally,  the bosonic flow equation  \eqref{flow_d1}
provides additional information on the system. 
The right hand side of the flow equation  vanishes 
if one uses the partition function of the form \eqref{lnZ_final}.
Therefore, $d_1$ should vanish. 
On the other hand one can obtain $d_1$ using the spectral curve
 \eqref{spectral_curve-B}.
It is noted in 
\cite{MMM_2011, BMT_2011b} 
that the resolvent at NS limit is of the form 
\be
\omega_B(z) = \epsilon ( \ln P(z))' 
= \epsilon  \sum_{\alpha=1}^N \frac1{z-z_\alpha} 
\label{omegaB_poles}
\ee  
with a monic polynomial of degree $N$ 
\be
P(z) = \prod_{\alpha=1}^N (z-z_\alpha) = \sum_{k=0}^N  p_{N-k} z^k 
\ee 
 with $p_0=1$.  
Then,  \eqref{spectral_curve-B} results in   
\be
d_1= \epsilon N  \left(  c_1  + p_1 (\epsilon  (N-1)/2 +  c_0)  \right). 
\label{d_1}
\ee
Note that $p_1$ is the sum of all the poles $p_1=\sum_\alpha z_\alpha$ 
and is same as the expectation value  $\langle \sum_I z_I \rangle$
of the matrix model. 
Since $d_1$ vanishes, one concludes that 
\be
p_1 = -   \frac{ c_1 }{  c_0+ \epsilon  (N-1)/2  }.  
\label{P_1}
\ee
The result \eqref{P_1} is consistent with constraint
$\ell_0 (d_1 ) = d_1$ since $  \ell_1 (d_0)=0$.
In general, there are two  variables $c_1$ and $\xi_0\xi_1/c_1$ 
of conformal dimension 1 for the rank 1 case. 
However, the term proportional to $\xi_0\xi_1/c_1$  turns out to vanish
and the only term proportional to $c_1$  survives. 

The pole structure of the bosonic resolvent also shares with that
of the fermionic one.
This can be seen from  \eqref{spectral_curve-F}.
First, note that  if one uses \eqref{omegaB_poles},
one may put  
 $x_B(z) = \epsilon ( \ln  \widetilde{P}_N(z))' $  with 
$\widetilde{P}_N = {P}_N e^{V_B/\epsilon}$.
Then,  \eqref{spectral_curve-F} reduces to 
\be
\epsilon  \widetilde{P}_N '(z) x_F(z) +  \epsilon \widetilde{P}_N(z)x_F'(z) ) =  \widetilde{P}_N(z) F_F(z) 
\ee
or 
$(\epsilon  \widetilde{P}_N (z) x_F(z) )' = \widetilde{P}_N(z) F_F(z) $. 
Therefore, $x_F(z)$ has the simple expression 
\be
x_F(z) =  \frac{\tau_F(z) } { {P}_N(z)}
\ee
where $\tau_F(z)=e^{- V_B(z) /\epsilon} \int^z dy  F_F(y) \widetilde{P}_N(y) /\epsilon  $.
Since $\tau_F(z_\alpha)$ is not zero in general (except 0 accidentally), 
the obvious conclusion is that the pole position $z_\alpha$ 
is also the pole position of $x_F(z)$.

\section{Partition function of half-odd rank}
\label{sec_4}

The partition function of integer rank $n$ is interpreted as the inner-product 
between a primary state and an irregular state of rank $n$.  
The spectral curve shows that the irregular state is the simultaneous  eigenstate of 
super-current $G_r$ with $r=n+1/2, \cdots, 2n-1/2$ 
and Virasoro current $L_m$ with $m= n, \cdots, 2n$ if $d_n=0$
(otherwise, $m=n+1,  \cdots, 2n$). 
One may wonder if the  eigenvalue of the highest Virasoro mode $L_{2n}$ 
vanishes. 

Note that the eigenvalue of the  highest Virasoro mode  is given as 
$\Lambda_{2n}= c_n^2$. Therefore, unless $c_n=0$ 
the case is not achieved in the NS sector. 
Instead, if one includes the Ramond sector also, one may have 
the potential of half-odd rank as in \eqref{ramond_V_B} and \eqref{ramond_V_F}.
\be 
V_B(z_I)   = c_0  \ln  (z_I) - \sum_{k =1}^{n} \frac{c_k} {k  z_I ^k} , ~~
V_F(z_I)  =  - \sum_{k =0}^{n-1} \frac{\xi_k}{z_I^{k+1}} .
\label{ramond_potential}
\ee
The difference from the NS sector is that the commuting variable 
vanishes when squared  $ c_n^2=0$. 
This is because $c_n$ is
commuting but is  the product of two anti-commuting variables. 
Therefore, the eigenvalue $\Lambda_{2n}$  vanishes 
so that the non-vanishing highest mode becomes $L_{2n-1}$.

We will consider two simplest cases: rank 1/2 and 3/2. 
The rank $1/2$ has bosonic parameters $ c_0, c_1$ and one fermionic  $\xi_0$
with the constraint $c_1^2=0=c_1 \xi_0$. 
It is clear that $\Lambda_n=0$ when $n\ge 2$ and  $\Lambda_1= c_1 (c_0 - \epsilon)$
so that the irregular state has the eigenvalue of highest Virasoro mode $L_1$. 
Note that $G_{3/2}$ annihilates the irregular state 
since $\Omega_{3/2} = c_1 \xi_0 =0$. 

The super-flow equations are simply given as 
\begin{align}
d_0 = & \left( c_1 \frac{\partial}{\partial c_1 }
 + \frac12  \xi_0 \frac{\partial}{\partial \xi_0 } \right)  (-\hbar^2 \ln Z)
\label{rank_1/2_flow_d0}\\
\eta_{1/2} =&  \left(  -c_1  \frac{\partial}{\partial \xi_0 } \right)
(-\hbar^2 \ln Z) + \epsilon \xi_0.
\label{rank_1/2_flow_eta1} 
\end{align}
The partition function is formally given as 
\be
-\hbar^2 \ln Z
= d_0  \ln c_1
\ee
where $d_0 =  \epsilon  N \left(c_0 + \frac{\epsilon (N-1)}2  \right)$ 
as given in \eqref{d_0}. 

Non-trivial case starts with rank 3/2. 
In this case there are three commuting parameters $c_0, c_1, c_2$ and 
two anti-commuting parameters $\xi_0, \xi_1$. 
The parameters have the relation with the original  $\gamma_k$ in \eqref{ramond_vertex}
as follows: 
$c_0 =\hbar \gamma_0$, 
$c_1=\hbar(\gamma_1 \theta+ \gamma_2)$, 
$c_2 = \hbar\gamma_3 \theta$, 
$\xi_0 = \hbar \gamma_1$,
and $\xi_1= \hbar(\gamma_2 \theta + \gamma_3)$. 
This shows that $c_2^2= c_2 \xi_1=0$.  

Then we have 4 flow equations
\begin{align}
d_0 = & \left( c_1 \frac{\partial}{\partial c_1 } 
+2 c_2 \frac{\partial}{\partial c_2 } 
+ \frac12  \xi_0 \frac{\partial}{\partial \xi_0 }
+\frac32  \xi_1 \frac{\partial}{\partial \xi_1 } \right)  (-\hbar^2 \ln Z)
\label{ramond_3/2_flow_d0}\\
\eta_{1/2} =&  \left( \xi_1 \frac{\partial}{\partial c_1 }  -c_1  \frac{\partial}{\partial \xi_0 } 
 -c_2  \frac{\partial}{\partial \xi_1 } \right)
(-\hbar^2 \ln Z) + \epsilon \xi_0
\label{ramond_3/2_flow_eta1} \\
d_1  = & \left ( c_2 \frac{\partial}{\partial c_1 } 
+ \xi_1  \frac{\partial}{\partial \xi_0  } \right) (-\hbar^2 \ln Z).
\label{ramond_3/2_flow_d1} \\
\eta_{3/2} =&  \left(   -c_2  \frac{\partial}{\partial \xi_0 }  \right)
(-\hbar^2 \ln Z)
\label{ramond_3/2_flow_eta3} 
\end{align}

The bosonic spectral curve \eqref{spectral_curve-B} shows that
$d_0$ is the same as in \eqref{d_0} and the solution of \eqref{ramond_3/2_flow_d0}
is given as 
\be
-\hbar^2 \ln Z = d_0 \log c_1 + A(t)\xi_0 \xi_1/c_1^2 + B(t)
\label{ramond_lnZ_try1}
\ee
where we use the fact 
 $t= c_2/c_1^2$ and $ \xi_0 \xi_1/c_1^2$ are homogeneous solutions.

The  fermionic spectral curve \eqref{spectral_curve-F} shows that 
$\eta_{1/2}$ has the same form \eqref{eta_1/2_try1}. 
However, the right hand side of the fermionic flow equation 
 \eqref{ramond_3/2_flow_eta1} has a different result. 
\begin{align}
\eta_{1/2}= &\xi_0 \Big( \epsilon  N 
+ \epsilon N_1   (c_0 + \epsilon (N-1)  ) \Big)
+\left( \frac{\xi_1} {c_1} \right) \Big( c_0 +N_2\epsilon (N-1) \Big) 
\nonumber \\
= &\xi_0 ( \epsilon+  t A( t) ) 
+ \left(  \frac{ \xi_1}{c_1} \right) ( d_0 - A(t) - 2t B'(t) ). 
\end{align}
 This fermionic flow equation reduces to another algebraic identity 
whose solves $A(t)$ and $B(t)$:
\be
 A(t) =   \frac{A_1  }t, ~~ B(t) = B_0 \log t +   \frac {B_1}t.
\ee
so that the partition function is given as 
\be
-\hbar^2 \ln Z = (d_0-2B_0)  \log c_1
+B_0 \log c_2
 +A_1   \xi_0 \xi_1/c_2 
 +   {B_1c_1^2}/{c_2} 
\ee
where $A_1=  \epsilon(  N-1  + N_1 ( c_0+ \epsilon (N-1) )$,
$B_0=    (d_0-  c_0 + \epsilon N_2  (N-1)  ))/2 $
and 
$B_1= \epsilon  (  N-1  + N_1 ( c_0+ \epsilon (N-1) )/2$. 
Therefore, the partition function \eqref{ramond_lnZ_try1} is given 
in terms of potential variables together with $N$, $N_1$ and $N_2$. 

Two more flow equations provide additional information on the system. 
$d_1$ is given as \eqref{d_1} and corresponding flow equation \eqref{ramond_3/2_flow_d1} 
shows that 
\be
\epsilon N  \left(  c_1  + p_1 (  c_0+ \epsilon  (N-1)/2 )  \right)
=(d_0 -2 B_0) c_2/c_1 + 2 B_1 c_1.
\ee
This gives the information on $p_1 =\langle \sum_I z_I \rangle$;
\be
p_1= \frac{(d_0 -2B_0)c_2/c_1 + (2 B_1 -N)c_1}
{\epsilon N (  c_0+ \epsilon  (N-1)/2 ) }.
\ee

Finally, the fermionic flow equation  \eqref{ramond_3/2_flow_eta3}   
shows that the right side is given as    
\be
RHS  = -A_1 \xi_1.
\ee 
On the other hand, fermionic spectral curve \eqref{spectral_curve-F} 
shows that the left hand side is 
\be
LHS = \epsilon( c_0  +\epsilon (N -2)) q_1
+ \epsilon( \xi_0 p_1  +\xi_1 N + (c_1  + \epsilon  p_1) N_F ) 
\ee
where $q_1= \langle \sum_I z_I \theta_I \rangle$,
fermionic partner of $p_1$. 
Therefore, the flow equation determines $q_1$. 
\be
q_1 = 
 - \frac{ A_1 \xi_1 + \epsilon( \xi_0 p_1  +\xi_1 N + (c_1  +\epsilon  p_1) N_F ) }
{ \epsilon( c_0  +\epsilon (N -2))} 
\ee
 
Note that  $\Lambda_n=0$ when $n\ge 4$ and 
 the positive Virasoro 
generators $L_3$ and $L_2$ have non-vanishing eigenvalues  $\Lambda_3= c_1 c_2 $ and 
$\Lambda_2 =c_1^2/2 +  (c_0 - 3\epsilon/2) c_2$, 
respectively. 
In addition, super-current $G_{n-1/2}$ with $n\ge 4$
annihilates the state and $G_{5/2}$
have non-vanishing eigenvalue 
 $\Omega_{5/2} = c_1 \xi_1 + c_2 \xi_0$.
This eigenvalue is consistent with the commutation algebra 
$G_{5/2}^2 = - L_{5}$ since $\Omega_{5/2}^2 = 0$
and $\Lambda_5 =0$. 

\section{Irregular Vertex Operators and RG Flow Equations}
In this section, we provide RG flow equations to the operator algebra
of the irregular vertices from the string field theory.
The main idea is that, in the formalism of irregular vertex operators,
we may have conformal $\beta$-function equations
on the wavefunctions of these operators, generalized to the off-shell case. 
For simplicity, we shall limit ourselves to the non-supersymmetric case
and to the rank one, however, the discussion is straightforward to 
generalize to higher ranks and the supersymmetry.
The most general form of the rank 1 vertex operator is
given by 
\begin{eqnarray}
U(\alpha,\beta)=\xi(\alpha,\beta)e^{\alpha\phi+\beta\partial{\phi}}
\end{eqnarray}
where $\xi(\alpha,\beta)$ is the wavefunction for the irregular state.
In case if $U(\alpha,\beta)$ were a regular vertex operator,  its leading order 
contribution to the string sigma-model partition function would be given by
$Z_{\sigma}\sim{e^{S(\xi)}}$ where $S(\xi)$ is the low-energy effective action, defined by the
vanishing $\beta$-function condition
\begin{eqnarray}
{{\delta{S}}\over{\delta{\xi}}}=\Lambda{{d\xi}\over{d\Lambda}}\equiv\beta_\xi\sim
\Delta{\xi}+C\xi^2 + O(\xi^3)=0
\end{eqnarray}
where $\Lambda$ is the worldsheet cutoff and $C$ are the structure constants defined by
3-point worldsheet correlators.
The above condition ensures that the conformal invariance is preserved
by inserting the on-shell operators on the worldsheet.
The irregular vertex operators are, however, the off-shell objects, therefore
they do not have any associate $\beta$-function in a naive literal sense. 
Nevertheless, the relation of the type (5.2)  still retains some important meaning off-shell,
in particular, in the context of background-independent string field theory - and can be related
to the flow equations derived above.
That is, in the on-shell case, the equations of motion (5.2) define the perturbative background
deformations preserving the worldsheet conformal symmetry, ensured by the Weyl invariance combined
along with the condition of absence of logarithmic singularities
in the partition function due to collisions between vertex operators. It is furthermore important that,
in the on-shell case, all the vertex operators have conformal dimension 1, and the only OPE terms
contributing to the $\beta$-functions as a result of collision of two such vertices, are those involving
operators of dimension one. In the off-shell case, such as ours,
 all these conditions have to be modified.
First of all,  $U(\alpha,\beta)$ becomes a string field which wavefunction, 
$\xi(\alpha,\beta)$ now describes a
$nonperturbative$ background deformation 
from the original to the one defined by the appropriate
analytic solution in string field theory.
The ``$\beta$-function''-like constraint of the type (5.2) is now precisely the condition that the string field $U$
is that analytic solution, producing the nonperturbative background change. Moreover, 
contrary to the perturbative on-shell-case, the ``effective action'' $S(\xi)$ is typically nonlocal.

For the irregular vertices, we can no longer require  the absence of the OPE singularities
for the colliding operators, as this constraint has an essentially on-shell origin in string perturbation theory.
However, we still have to retain the Weyl invariance constraints on the operators,
since a) these constraints are imposed off-shell even in standard string perturbation theory b) Weyl invariance is essential to fix the (super)conformal gauge
which we are using here.
In order to elucidate the constraints due to the scale invariance, one
has to calculate the OPE of the irregular vertex sitting on the disc boundary,
with the trace of the stress-energy tensor, integrated over the bulk of the disc,
 and to extract the logarithmic divergence stemming from the OPE integration.
This shall lead to the first set of the constraints, analogous to the flow equations.
Straightforward calculation gives:
\begin{align}
& \lim_{z,{\bar{z}}\rightarrow{\tau}}
:T_{z{\bar{z}}}:(z,{\bar{z}}):e^{\alpha\phi+\beta\partial\phi}:(\tau)
= 
\lim_{z,{\bar{z}}\rightarrow{\tau}}
-{1\over2}:\partial\phi{\bar{\partial}}\phi:(z,{\bar{z}})
e^{\alpha\phi+\beta\partial\phi}(\tau)
\nonumber \\
&~~~=
\left\lbrace
{\alpha^2}\over{{2|z-\tau|^2}}
\right.+
\left(
{{({{z-{\bar{z}}}})^2}\over{|z-\tau|^4}}+{{2}\over{|z-\tau|^2}} \right)
:{{\beta}}\partial{{\phi}}e^{\alpha\phi+\beta\partial\phi}:(\tau)
\nonumber \\
&
~~~~~~~~~
+{1\over{|z-\tau|^2}}\left\lbrack
{{\beta^2}\over8}({{\alpha}}\partial{{\phi}})^2
-{{\alpha}}\partial^2{{\phi}}
+2
({{\alpha}}\partial{{\phi}})({{\alpha}}\partial^2{{\phi}})
-{{\beta}}\partial^3{{\phi}}\right\rbrack
\nonumber \\
&
~~~~~~~~~~~~
\left.
-{1\over2}{{{{\alpha}}{{\beta}}}\over{|z-\tau|^2}}
({{\alpha}}\partial{{\phi}}+{{\beta}}\partial^2{{\phi}})
\right\rbrace
{e^{\alpha\phi+\beta\partial\phi}}:(\tau)
\end{align}
Integrating over $z$ the contributions proportional to 
$\sim\int{d^2z}{1\over{|z-\tau|^2}}\sim{\ln\Lambda}$ leads to logarithmic singularities
defining the variations of the operators under Weyl transformations.
Cancellation condition for these variations  defines the flow equations  we are
looking for. In what follows we shall ignore the OPE terms with
higher derivatives of $\phi$. That is, the terms proportional to
$\partial^2\phi$ and higher derivatives, are 
only relevant for the RG flows for the 
 higher rank operators, related to variational
derivatives  with respect momenta, conjugate to higher derivatives in the 
irregular vertices
(e.g. $:\partial^2\phi{e^{\alpha\phi+\beta\partial\phi+\gamma\partial^2\phi}}:
\sim{{\partial}\over{\partial\gamma}}
{e^{\alpha\phi+\beta\partial\phi+\gamma\partial^2\phi}}$)
Then the flow equation describing the Weyl deformations of 
the irregular operators is
\begin{eqnarray}
\beta_\xi=\Lambda{{d\xi}\over{d\Lambda}}=
-{{\alpha^2}\over2}\xi-{{{\beta}}}{{\partial}\over{\partial{{\beta}}}}\xi
-{{\beta^2}\over{8}}({{{\alpha}}}{{\partial}\over{\partial{{\beta}}}})^2\xi
-{1\over2}({{\alpha}}{{\beta}})
{{\alpha}}
{{\partial}\over{\partial{{\beta}}}}\xi
\end{eqnarray}

This extended $\beta$-function relation 
is related to the Legendre transformed bosonic part of the flow equation 
(3.4) for the free energy $\ln{Z}$, 
expressed in terms of the wavefunction $\xi(\alpha,\beta)$, related to the partition function according to
\be 
Z_{\sigma}=\sum_P{1\over{P!}}\int{d\tau_1}...{d\tau_P}\xi(\alpha_1,\beta_1)
...\xi(\alpha_P,\beta_P)<V(\alpha_1,\beta_1,\tau_1)...V(\alpha_P,\beta_P,
\tau_P)>
\ee
where $V(\alpha,\beta,\tau)=:e^{\alpha\phi+\beta\partial\phi}:(\tau)$.

The relation to the bosonic part of the flow equation (3.4) 
 is  not straightforward
because  the generalized RG flow (5.4) is expressed in terms of very 
different variables.  To obtain this relation, one has to insert
the differential operator on the right hand side of (5.4) inside
the generating functional 
$<e^{\int{d\tau}d\alpha{d\beta}\xi(\tau,\alpha)V(\alpha,\beta,\beta)}>$.
The relation will then follow as the $2d$ Ward identity inside the 
worldsheet correlators.

It is straightforward to generalize this calculation to the supersymmetric 
case.
In this case, the irregular vertices are not eigenvalues of positive
Virasoro generators, but the Jordan blocks. In the simplest rank ${1\over2}$
case such a block has a multiplicity 2 with components:
\begin{eqnarray}
V_1=\eta_1(\alpha,\beta)(\alpha\psi+\beta\partial\phi)
\nonumber \\
V_2=\eta_2(\alpha,\beta)e^\phi
\end{eqnarray}
Applying the Weyl transformation now leads to separate equations
on the wavefunctions $\eta_1$ and $\eta_2$:
\begin{eqnarray}
(\alpha^2+{{\beta}}{{\partial}\over{\partial{{\alpha}}}})\eta_1
+{{\alpha}}{{\beta}}{{\alpha}}
{{\partial}\over{\partial{{\alpha}}}}\eta_1=0
\nonumber \\
(\alpha^2-1)\eta_2=0
\end{eqnarray}
Note that, unlike (4.3) in the pair of the flow equations, 
one of the equations for the rank ${1\over2}$ is algebraic.

\section{Conclusion and discussion}
In this work, we analyzed the loop equation in supersymmetric matrix model
in the superspace formalism,
in order to derive the spectral curve for the 
Argyres-Douglas limit of $N=2$ super Yang-Mills theory, related to N=1 super Liouville
conformal field theory through generalized AGT conjecture.
We have been able to derive and to integrate the loop equation in the supersymmetric case
and to obtain partition functions  associated with irregular blocks
of ranks ${1\over2}$, 1 and ${3\over2}$.

The loop equations, as well as the associate flow equations on the free energy,
can be reproduced in the irregular vertex operator approach, in terms of the scale invariance 
constrants for the vertex operators. 
One particularly promising thing about the vertex operator 
approach is that it is relatively straightforward
to extend to higher ranks, 
as well as to observe the Jordan cell structure of the flow equations in the 
supersymmetric case. 
 We hope to be able to extend these results 
to higher/arbitrary ranks in the future works.
It will be also interesting to investigate  (super)-spectral curve 
for the special value of Liouville parameter space
as observed in \cite{manabe_2016}. 

In general, it is natural to understand the AGT conjecture as an isomorphism between the partition
functions of the sigma-models with irregular vertex operators in Toda/superstring theories
and those of super Yang-Mills theories. The relation between these theories can be thought of as a generalization
of the one between standard string-theoretic sigma-models and their low-energy limit, through the off-shell
generalization of the conformal $\beta$-functions. The background-independent second-quantized string field theory
approach appears to be a promising framework for that.   The work in this direction is currently in progress
and we hope to be able to elaborate on these issues soon.

\begin{center}
{\bf Acknowledgements}
\end{center}
We  thank M. Manabe for discussion on the works \cite{manabe_2016}. 
The authors acknowledge the support of this work by the National Research Foundation of Korea(NRF)
 grant funded by the Korea government(MSIP) (NRF-2014R1A2A2A01004951) and by the National Natural 
Science Foundation of China under grant 11575119.

\appendix
\section{Super-spectral curve} 
\label{append_A}
One may derive the loop equation of the irregular super-matrix model 
corresponding to the super-conformal symmetry. 
Spin 3/2 current contribution is obtained if one use the supercoordinate transform
\cite{manabe_2016}
$z_I \to z_I + \theta_I \epsilon_F /(z-z_I)$ 
and $\theta_I \to \theta_I + \epsilon_F/(z-z_I)$ 
where $\epsilon_F$ is the small anti-commuting number.  
The metric contribution 
\be
\left[\prod_I dz_I d\theta_I \right] 
\to
\left[\prod_I dz_I d\theta_I \right] 
\left( 1 + \sum_I \frac{\theta_I ~\epsilon_F}{(z-z_I)^2} \right) . 
\ee
Super-Vandermonde determinant has the contribution
\be
\prod_{I<J} (z_{IJ} - \theta_I \theta_J )^{\beta}
\to 
\prod_{I<J} (z_{IJ} - \theta_I \theta_J )^{\beta}
\left( 1+ \beta\left\{  \sum_{I,J} \frac{\theta_I}{(z-z_I)(z-z_J)}  -\sum \frac{\theta_I}{(z-z_I)^2}
\right\}\epsilon_F  \right).
\ee
Finally, the potential has the contribution
\be
e^{\frac{\sqrt{\beta} } g   \sum_{I}  V(\zeta_I)} 
\to 
e^{\frac{\sqrt{\beta} } g   \sum_{I}  V(\zeta_I)}
\left(1 +  \frac{\sqrt{\beta} } g
\sum_I 
\left\{\frac{ V_B'(z_I) \theta_I -V_F (z_I) }{z-z_I} \right\}
\epsilon_F
\right).
\ee
Collecting all terms one has 
\be
\omega_B(z) \omega_F(z) + V_B'(z) \omega_F(z) 
-V_F(z) \omega_B  - \hbar^2 \omega_{BF}(z,z) + \hbar b \omega_F'(z) 
= f_F(z) 
\label{loop_F}
\ee
where prime denotes the derivative with respect to $z$. 
$\omega_B(z)$ ($\omega_F(z)$) is  one-point commuting (anti-commuting) resolvent
\be
\omega_B (z) = g \sqrt{\beta} \left\langle \sum_I \frac1{z-z_I} \right\rangle,~~~
\omega_F (z) = g \sqrt{\beta} \left\langle \sum_I \frac {\theta_I}{z-z_I} \right\rangle.
\ee 
$\omega_{BF}(z,z) $ is the connected two-point resolvent
\be
\omega_{BF}(z,w) = {\beta} \left\langle \sum_I \frac1{z-z_I}   \sum_J \frac {\theta_I}{w-z_J}  \right\rangle_{ \!\!\! \rm conn}.
\ee
$f_F$ is related with the super-potential
\be 
f_F(z)  \equiv  g  \sqrt{\beta} 
 \left\langle 
\frac{( V_B'(z) -V_B'(z_I) ) \theta_I - (V_F(z) -V_F(z_I)) } {z-z_I}.
 \right\rangle 
\ee

Virasoro contribution is obtained if one uses the 
super-coordinate transform
$z_I \to z_I +\epsilon /(z-z_I) $ and 
$\theta_I \to \theta_I (1 + \epsilon/(2 (z-z_I)^2 ) $
where  $\epsilon$ is an infinitesimal commuting number. 
The metric contribution is 
\be
\left[\prod_I dz_I d\theta_I \right] 
\to
\left[\prod_I dz_I d\theta_I \right] 
\left( 1 +  \frac \epsilon2  \sum_I \frac{1}{(z-z_I)^2} \right) . 
\ee
(Here the anti-commuting measure $[d\theta_I ]$  is required
 to maintain the integral property 
$\int d\theta_I \theta_I =1$). 
Super-Vandermonde determinant has the contribution
\begin{align}
& \prod_{I<J} (z_{IJ} - \theta_I \theta_J )^{\beta}  
\to 
\prod_{I<J} (z_{IJ} - \theta_I \theta_J )^{\beta}\\
&~~~~~
\times
\left( 1+ \epsilon    \frac{ \beta  }2  \left\{  \sum_{I,J} 
\left[
\frac{1 }{(z-z_I)(z-z_J)} 
+\frac{\theta_I}{z-z_I} \frac{\theta_J}{(z-z_J)^2} 
\right] 
 -\sum_I \frac{1}{(z-z_I)^2} 
\right\} \right).
\end{align}
Finally, the potential has the contribution
\be
e^{\frac{\sqrt{\beta} } g   \sum_{I}  V(\zeta_I)} 
\to 
e^{\frac{\sqrt{\beta} } g   \sum_{I}  V(\zeta_I)}
\left(1 + \epsilon  \frac{\sqrt{\beta} } g
\sum_I  
\left[ 
\frac{( V_B'(z_I)  + \theta_I  V_F'(z_I) ) }{z-z_I}
+ \frac{\theta_I V_F(z_I)}  {2 (z-z_I)^2}
\right] 
\right).
\ee
Collecting all terms one has 
\begin{align}
\frac12 
\omega_B(z)^2+ V_B'(z) \omega_B(z)  
+\frac12 \Big(\omega_F(z) V_F'(z) - \omega_F'(z) V_F(z) \Big) 
\\
 + \frac{\hbar Q}2  \omega_B'(z) 
+ \frac12 \hbar^2 \left(\omega_{BB}(z,z) + \omega_{FF}^{(1,2)}(z,z) \right)
 = f_B(z) 
\label{loop_B}
\end{align}
where 
$\omega_{BB}(z,z) $ and $\omega_{FF}^{(1,2)}(z,z)$ are 
 the connected two-point resolvents
\be
\omega_{BB}(z,w) 
= {\beta} \left\langle \sum_I \frac1{z-z_I}   \sum_J \frac { 1}{w-z_J}  \right\rangle_{ \!\!\! \rm conn}.
\ee
and 
\be
\omega_{FF}^{(1,2)}(z,w) 
= {\beta} \left\langle \sum_I \frac{\theta_I}{z-z_I}  
 \sum_J \frac { \theta_J}{ (w-z_J)^2}
  \right\rangle_{ \!\!\! \rm conn}.
\ee
$f_B$ is related with the super-potential  
\be 
f_B(z) = g \sqrt{\beta} \left\langle 
\sum_I  \frac{(V_B'(z) -V_B'(z_I)) + \theta_I (V_F'(z) -V_F'(z_I))  }{z-z_I}  +\frac12 \frac{ \theta_I (V_F (z) -V_F (z_I))  } {(z-z_I)^2} \right\rangle. 
\ee

It is useful to find the explicit holomorphic structure of $f_F(z)$ and $f_B(z)$
for the given potential \eqref{V_B} and \eqref{V_F}.
They are given in terms of the inverse powers of $z$ 
\be
f_F(z)= \sum_{r=-1/2}^{n-1/2} \frac{\eta_{r}} {z^{3/2+r}}.
\label{f_F}
\ee
The moment $\eta_r$ is given as an expectation value
and $\eta_{-1/2}$ vanishes which is evident 
from $1/z$ expansion of  \eqref{loop_F}. 
If one uses the explicit form of the potential, one may 
put the non-vanishing  moment into an 
interesting form as in non-supersymmetric case 
\cite{NR_2012, CR_2015} 
\be 
 \eta_r = g_r (- \hbar^2 \log Z) + \delta_{r, 1/2} g \sqrt{\beta} \xi_0 
\label{etar_gr}
\ee
where $g_r$ is the 
differential representation of the super current 
(corresponding to right action)
\be
g_r  = \sum_k   
\left(  k \xi_{k + r -1/2} \frac{\partial} {\partial c_k}
- c_{k + r +1/2} \frac \partial {\partial \xi_k} \right).
\label{g_r}
\ee
This is obtained if one notices that 
\be
\frac{\sqrt \beta}{g} 
\left \langle \frac{1}{z_I^{k+1}} \right \rangle =  k\frac{\partial }{\partial c_{k} } \ln Z , 
~~~
\frac{\sqrt \beta}{g} 
\left \langle \frac{\theta_I}{z_I^k} \right \rangle =  \frac{\partial }{\partial \xi_k } \ln Z 
\ee

Likewise, $f_B$ is written in terms of inverse powers of $z$, 
\be
f_B(z) = \sum_{m=-1}^{n} \frac{ d_m  }{z^{2+m} }. 
\label{f_B}
\ee
The moment $d_{-1}$ vanishes from $1/z$ expansion of  \eqref{loop_F}. 
Non-vanishing  moment has the form  
\be
 d_m = \ell_m  (- \hbar^2 \log Z)
\label{dm_ellm}
\ee  
where $\ell_m$ is the 
differential representation of the Virasoro current 
(corresponding to right action)
\be
\ell_m  =\sum_k  \left (  l\, c_{l+m}  \frac{\partial}{\partial c_l} 
+\left( \frac{ 2\ell+ m+ 1}2 \right) \xi_{l +m} \frac{\partial}{\partial \xi_l} \,.
\right)
\label{ell_m}
\ee
It can be checked  that $g_r$ in \eqref{g_r} 
and   $l_m$ in \eqref{ell_m} satisfy 
the commutation relation of right action
of the super algebra
\be 
[l_m, g_r]= (r - \frac{m}2 ) g_{r+m}, ~~
\{g_r, g_s \} = -2 l_{r+s},~~
  [ l_m , l_n ] = -(m-n) l_{m+n}. 
\label{super_alg}
\ee

At the NS  limit 
($\hbar \to 0$ and $b \to \infty$ so that $\hbar b = \epsilon $), 
the loop equations \eqref{loop_F} and \eqref{loop_B} 
can be put in terms of one-point resolvent only,
which is called the deformed spectral curve
\begin{align} 
& x_B (z) x_F (z) + \epsilon  x_F '(z)  = F_F (z) 
\\
&x_B(z)^2 + \epsilon x_B'(z) + x_F(z) V_F' (z) - x_F'(z)  V_F(z) = 2F_B (z)   
\end{align}
where we use compact notations: 
$ x_B(z) = \omega_B(z) + V_B'(z)$,  
$x_F(z) = \omega_B(z) - V_F(z) $, 
$  F_F(z) = f_F(z) -V_B'(z) V_F(z) - \epsilon V_F' (z)$ and  
$ F_B(z) =f_B(z) + \frac12 V_B'^2  + \epsilon V_B'(z)$.

It is interesting to look into the explicit form of $F_F (z) $ and $F_B(z) $.   
\be
F_F (z) = \sum_{r=1/2}^{2n+1/2} \frac{\Omega_r + \eta_r }{z^{3/2+r} },
~~~
F_B (z) = \sum_{m=0}^{2n} \frac{\Lambda_m + d_m  }{z^{2+r} },
\ee
where $\Omega_r$ is  an anti-commuting number
$\Omega_r = \sum_{k+\ell= r -1/2}  c_k \xi_{\ell} - \epsilon ( \delta_{r, 1/2} -(r+1/2)) \xi_{r-1/2} $
and $\Lambda_m$ is a commuting number
$\Lambda_m =  \sum_{k+l =m} c_k c_l/2   - \epsilon (m+1) c_m /2 $ 
Non-vanishing 
$\eta_r$ ($r=1/2, \cdots, n-1/2$)
 and $d_m$ ($m=0, \cdots, n$) 
are  given in \eqref{f_F} and  \eqref{f_B}.
The anti-commuting number 
$\Omega_r $  with $r=( n+1/2, n+3/2, \cdots, 2n+1/2)$
corresponds to the eigenvalue of super-current positive mode 
$G_r$
and the commuting number  $\Lambda_m$
with  $m=(n+1, n+2, \cdots, 2n)$
corresponds to the eigenvalue of Virasoro positive 
mode $L_m$. 

The same analysis can be done for the potential 
\eqref{ramond_V_F}
and \eqref{ramond_V_B}
of the  half-odd rank ($n-1/2$) similarly 
if one considers the constraint of the variables,
$c_n^2 =0=c_n \xi_{n-1}$. 
This shows that 
$\Lambda_m=0 $ if $m \ge 2n$ 
and $\Omega_r =0$ if $r \ge 2n-1/2$. 
 


\begin{thebibliography}{99}

\bibitem{G_2009}
D. Gaiotto,  \emph{Asymptotically free N=2 theories and irregular conformal blocks},
J.Phys. Conf. Ser. {\bf 462} (2013) no.1, 012014 [arXiv:0908.0307].

\bibitem{Whittaker}
E. Felinska, Z. Jaskolski, and M. Kosztolowicz, \emph{Whittaker Pairs for the Virasoro Algebra and
the Gaiotto - Bmt States},   Math.\ Phys.\   {\bf 53} (2012) 033504 [arXiv:1112.4453 [math-ph]].

\bibitem{MMM_2009}
A. Marshakov, A. Mironov and  A. Morozov,   
\emph{On non-conformal limit of the AGT relations},
 Phys. Lett. {\bf B682} (2009) 125 [arXiv:0909.2052].

\bibitem{BMT_2011}
G. Bonelli, K. Maruyoshi and A. Tanzini,
 \emph{Wild Quiver Gauge Theories}, JHEP {\bf 1202}  (2012) 031
[arXiv:1112.1691].

\bibitem{KMST_2013} 
H.~Kanno, K.~Maruyoshi, S.~Shiba and M.~Taki,
\emph{$W_3$ irregular states and isolated N=2 superconformal field theories},
JHEP {\bf 1303}, 147 (2013)
[arXiv:1301.0721 [hep-th]].


\bibitem{CRZ_2014} 
S.-K.\  Choi, C.\ Rim and H.\  Zhang, 
 \emph{Virasoro irregular conformal block and beta deformed random matrix model},
 Phys.\ Lett.\ {\bf B742} (2015) 50-54.


\bibitem{CRZ_2015} 
S.-K.\  Choi, C.\ Rim and H.\  Zhang, 
 \emph{Irregular conformal block, spectral curve and flow equations},
JHEP {\bf 1603} (2016) 118.



\bibitem{AGT_2010}
  L.~F.~Alday, D.~Gaiotto and Y.~Tachikawa,
 \emph{Liouville Correlation Functions from Four-dimensional Gauge Theories},
 Lett. Math. Phys. {\bf 91} (2010) 167 [arXiv:0906.3219 [hep-th]].

\bibitem{EM_2009} 
T. Eguchi and K. Maruyoshi, \emph{Penner Type Matrix Model and Seiberg-Witten Theory}, {JHEP} 
 {\bf 1002}  (2010) 022 [arXiv: 0911.4797].

\bibitem{GT_2012}
D. Gaiotto and J. Teschner, 
\emph{Irregular singularities in Liouville theory}, {JHEP} {\bf 1212}  (2012) 050
[arXiv:1203.1052]. 


\bibitem{DV_2009} 
R. Dijkgraaf and C. Vafa,
\emph{Toda Theories, Matrix Models, Topological Strings, and N=2 Gauge Systems},
arXiv:0909.2453 [hep-th].


\bibitem{IO_2010} 
H.~Itoyama and T.~Oota,
\emph{Method of Generating q-Expansion Coefficients for Conformal Block and N=2 Nekrasov Function by beta-Deformed Matrix Model},
Nucl.\ Phys.\ B {\bf 838}, 298 (2010)
[arXiv:1003.2929 [hep-th]].


\bibitem{PR_IV_2016}
D.\  Polyakov  and C.\ Rim, 
 \emph{Irregular Vertex Operators for Irregular Conformal Blocks},
Phys.\ Rev.\  {\bf D93} (2016) 106002 [arXiv:1601.07756[hep-th]]

\bibitem{NS_2010}
H. Nagoya and J. Sun,
\emph{Confluent primary fields in the conformal field theory}
J. Phys. A {\bf 43} (2015) 465203 [arXiv:1002.2598 [math-ph]]

\bibitem{GF_2014}
J. Gomis, B. Le Floch,
\emph{M2-brane surface operators and gauge theory dualities in Toda}
[arXiv:1407.1852[hep-th]]

\bibitem{PR_2016}
D.\  Polyakov  and C.\ Rim, 
 \emph{Vertex Operators for Irregular Conformal Blocks: Supersymmetric Case}, 
arXiv:1604.08741[hep-th]. 


\bibitem{Rash_Stanish_1996} 
R.\ C.\  Rashkov and M.\ Stanishkov,
 \emph{Three point correlation functions in N=1 superLiouville theory}, 
Phys.\ Lett.\  {\bf B380} (1996) 49-58
[hep-th/9602148].

\bibitem{NR_2012} 
Takahiro Nishinaka  and  Chaiho Rim,
 \emph{Matrix models for irregular conformal blocks and Argyres-Douglas theories},
JHEP  {\bf 1210}  (2012) 138
[arXiv:1207.4480 [hep-th]]. 

\bibitem{CR_2013}
S.-K.\  Choi, C.\ Rim,
 \emph{Parametric dependence of irregular conformal block},
JHEP {\bf 1404} (2014) 106 [arXiv:1312.5535[hep-th]].

\bibitem{CR_2015} 
Sang Kwan Choi, Chaiho Rim,
 \emph{Irregular matrix model with W symmetry}, 
J.\ Phys.\ { \bf A49} (2016) 075201
[arXiv:1506.02421 [hep-th]]. 


\bibitem{MMM_2011}
A\. Marshakov, A.\ Mironov  and  A.\ Morozov,
 \emph{On AGT Relations with Surface Operator Insertion and Stationary Limit of Beta-Ensembles}, 
J.\ Geom.\ Phys.\ {\bf 61} (2011) 1203-1222 
[arXiv:1011.4491 [hep-th]]

\bibitem{BMT_2011b} 
G.\  Bonelli, K.\  Maruyoshi, A.\ Tanzini, 
 \emph{Quantum Hitchin Systems via beta-deformed Matrix Models}, 
arXiv:1104.4016 [hep-th].

\bibitem{manabe_2016} 
M.\ Manabe and P.\  Sułkowski,
 \emph{Quantum curves and conformal field theory},  arXiv:1512.05785[hep-th];
P.\ Ciosmak, L.\  Hadasz, M.\ Manabe and P.\  Su lkowski
 \emph{Super-quantum curves from super-eigenvalues models}, 
arXiv:1608.02596[hep-th]


\end{thebibliography}
\end{document}